\documentclass{PoS}
\usepackage{graphicx}
\usepackage[authoryear,round]{natbib}

\title{The impact of the SKA on Galactic Radioastronomy: continuum observations}

\ShortTitle{Galactic Radioastronomy}

\author{
\speaker{Grazia Umana}$^1$, 
Corrado Trigilio$^1$,
Luciano Cerrigone$^{2}$,
Riccardo Cesaroni$^{3}$,
Albert  A. Zijlstra$^{4}$,
Melvin Hoare$^{5}$,
Kerstin Weis$^{6}$,
Anthony  J. Beasley$^{7}$,
Dominik Bomans$^{6}$,
Greg Hallinan$^{8}$,
Sergio Molinari$^{9}$,
Russ Taylor$^{10}$,
Leonardo Testi$^{11}$,
Mark Thompson$^{12}$
\\ 
$^1$INAF OACT, Catania, Italy; 
$^2$JCentro de Astrobiologia (INTA-CSIC), Madrid, Spain;
$^3$INAF Osservatorio Astrofisico di Arcetri, Firenze Italy; 
$^4$Jodrell Bank Centre for Astrophysics, Manchester M13 9PL, UK;
$^5$University of Leeds,  Leeds, UK;
$^6$Astronomisches Institut, Ruhr-Universitaet, Bochum, Germany;
$^{7}$National Radio Astronomy Observatory, Charlottesville, USA;
$^8$Department of Astronomy, California Institute of Technology, Pasadena, CA 91125, USA;
$^9$INAF IAPS, Roma, Italy;
$^{10}$ University of Cape Town and The University of the Western Cape, SA;
$^{11}$  ESO, Karl Schwarzschild str. 2, D-85748 Garching, Germany;
$^{12}$University of Hertfordshire, UK

\\
E-mail: \email{Grazia.Umana@oact.inaf.it}
}

\abstract{The SKA will be a state of the art radiotelescope optimized for both large area surveys as well as  
for deep pointed observations. In this paper we  analyze the impact that the SKA will have on Galactic studies, starting from the immense
 legacy value of the all-sky survey proposed by the continuum SWG but also presenting some areas of Galactic Science that particularly benefit from SKA observations both surveys and pointed.
 
The planned all-sky survey will be characterized by  unique spatial resolution, sensitivity and survey speed, providing us with a wide-field atlas of the Galactic continuum emission.  Synergies with existing, current and planned radio Galactic Plane surveys will be discussed.
SKA will give the opportunity to create a sensitive catalog of discrete Galactic radio sources, most of them representing the interaction of stars at various stages of their evolution with the environment: complete census of all stage of HII regions evolution; complete census of late stages of stellar evolution such as  PNe and SNRs; detection of stellar winds, thermal jets, Symbiotic systems, Chemically Peculiar and dMe stars, active binary systems in both flaring and quiescent states. 
Coherent emission events like Cyclotron Maser in the magnetospheres of  different classes of stars can be detected. 
Pointed, deep observations will allow new insights into the physics of the coronae and plasma processes in active stellar systems and single stars, enabling the detection of flaring activity in larger stellar population for a better comprehension of the mechanism of energy release in the atmospheres of stars with 
different masses and age.
}

\FullConference{
Advancing Astrophysics with the Square Kilometre Array\\
June 8-13, 2014\\
Giardini Naxos, Italy}

\newcommand{\skipthis}[1]{}

\newcommand\apj{ApJ}
\def\kms {\ifmmode{{\rm ~km~s}^{-1}}\else{~km~s$^{-1}$}\fi}

\def\lsun {\ifmmode{{\rm ~L}_\odot}\else{~L$_\odot$}\fi}

\def\arcsec {$^{\prime \prime}$}

\newbox\grsign \setbox\grsign=\hbox{$>$} \newdimen\grdimen \grdimen=\ht\grsign
\newbox\simlessbox \newbox\simgreatbox
\setbox\simgreatbox=\hbox{\raise.5ex\hbox{$>$}\llap
 {\lower.5ex\hbox{$\sim$}}}\ht1=\grdimen\dp1=0pt
\setbox\simlessbox=\hbox{\raise.5ex\hbox{$<$}\llap
 {\lower.5ex\hbox{$\sim$}}}\ht2=\grdimen\dp2=0pt

\def\lsim{\mathrel{\rlap{\lower4pt\hbox{\hskip1pt$\sim$}}
    \raise1pt\hbox{$<$}}}                
\def\gsim{\mathrel{\rlap{\lower4pt\hbox{\hskip1pt$\sim$}}
    \raise1pt\hbox{$>$}}}                

\def\apj {{\it Ap.~J.}}
\def\apjl {{\it Ap.~J.\ (Letters)}}
\def\apjs {{\it Ap.~J.\ Suppl.}}
\def\aj {{\it A.~J.}}

\def\aap {{\it Astr.~Ap.}}

\def\araa {{\it Ann.\ Rev.\ Astr.\ Ap.}}

\def\mnras {{\it MNRAS}}

\def\nat {{\it Nature}}
\def\pasa {{\it PASA}}
\def\pasp {{\it PASP}}

\def\aj{AJ}                   
\def\araa{ARA\&A}             
\def\apj{ApJ}                 
\def\apjl{ApJ}                
\def\apjs{ApJS}               
\def\aap{A\&A}                
\def\aaps{A\&AS}              
\def\mnras{MNRAS}             

\def\pasp{PASP}               
\def\nat{Nature}              

\def\grtsim{\mathrel{\hbox{\rlap{\hbox{\lower2pt\hbox{$\sim$}}}\raise2pt\hbox{$>$}}}}
\def\lesssim{\mathrel{\hbox{\rlap{\hbox{\lower2pt\hbox{$\sim$}}}\raise2pt\hbox{$<$}}}}

\def\lsim{\,\lower2truept\hbox{${<\atop\hbox{\raise4truept\hbox{$\sim$}}}$}\,}
\def\gsim{\,\lower2truept\hbox{${> \atop\hbox{\raise4truept\hbox{$\sim$}}}$}\,}
\def\simlt{\mathrel{\rlap{\lower 3pt\hbox{$\sim$}}
        \raise 2.0pt\hbox{$<$}}}
\def\simgt{\mathrel{\rlap{\lower 3pt\hbox{$\sim$}}
        \raise 2.0pt\hbox{$>$}}}

\begin{document}


\section{The all-sky SKA1 survey}

The proposed SKA1 all sky continuum survey (SASS1: Norris et al.\ 2015),
covering  most  ($\sim  70 \%$) of the Galactic Plane with unprecedented
spatial resolution, sensitivity and survey speed at band 2 (650-1670 MHz),
will  provide us with a sensitive wide-field atlas of Galactic continuum
emission.  Up to now,  existing interferometric radio continuum surveys of
the Galactic Plane have been carried out either at high angular resolution,
but over a limited survey area, or over wider areas but  at low angular
resolution. This is the case of MAGPIS \citep{Helfand_2006} and CORNISH
surveys \citep{Hoare_2012}, covering  an area of  $\sim$ square degrees  at
an angular resolution of 1-6 \arcsec.  On the other side,  the
International Galactic Plane Survey \citep{McClure_2005, Taylor_2003} and
the 2nd Epoch Molonglo Galactic Plane Surveys
\citep[MGPS-2:][]{Murphy_2007} cover several hundred square degrees at a
typical resolution of $\sim$1 arcmin.

In the meantime, two major surveys including or aimed at mapping the Galactic Plane will be carried out by the two SKA precursors, namely:
the Evolutionary Map of the Universe  (EMU, Norris et al., 2011)  a deep ($ \sim 10 \mu$Jy/beam), almost full sky
(75$\%$), $ \sim 10^{\prime \prime}$ angular resolution  survey to be carried out at 1.4 GHz with the Australian SKA  Pathfinder (ASKAP);
and MeerGal  (PIs  M. Thompson and  S. Goedhart) a
deep ($\sim 30  \mu$Jy/beam) survey  of the Milky Way Galaxy  (140 square degrees) to be carried out at 14 GHz with 
MeerKAT at a sub-arcsec  ($\sim 0.8 ^{\prime \prime}$) angular resolution. 

We can anticipate that both surveys will provide us with a new view   of the radio Galactic Plane.
However, SASS1, due to the increase in sensitivity,  angular resolution and survey speed, will bridge the gap between the two
types of existing  radio surveys of the Galactic Plane and will improve the results from EMU and MeerGAL, detecting and cataloguing objects such as HII regions, supernova
remnants (SNRs), planetary nebulae (PNe) and radio stars, down to the $\mu$Jy level. 
Most of these sources represent the interaction of stars at various stages of evolution with their environment.
The known radio populations of  each of these types of objects are limited by a combination
of issues including the limited area covered by existing surveys, sensitivity, angular resolution  or biases against large scale structures introduced by
limited uv coverage snapshot surveys. Moreover, the lessons learnt from EMU will guide the SASS1  design in  identifying  issues arising from the complex 
continuum structure associated with the Galactic Plane and from the variable sources in the Galactic Plane.\\
However,  besides  the legacy value of SASS1, some areas of Galactic Science will  particularly benefit from SKA observations, in particular 
from those conducted at higher frequencies (band 4 and/or  5)  and at higher resolution (sub-arcsec). 
 We will present  some among the possible Galactic science goals that can be addressed  with the SKA1  and the full SKA.
In the following,  we assume the SKA1 and SKA performances as reported in
the SKA1 Baseline  Design  (Dewdney et al., 2013) 
and in the Performance Memo (Braun, 2014).\\

\section{Massive star formation}

A straightforward application of SKA will be the study of ionized regions
around massive, early-type stars. Thanks to its overwhelming sensitivity and
dynamic range SKA will be the ideal instrument to detect and resolve these
objects, whose size may vary from $10^3$~au, for hypercompact (HC) HII
regions, to 100~pc, for giant HII regions. This is equivalent to span a broad
range of ages, as HII regions are known to expand with time, which makes SKA
an excellent tool to investigate the evolution of these intriguing objects.

While the most extended (and hence old) stages can be (and have effectively
been) studied with the Very Large Array, the youngest sources have been so
far quite elusive. The HC~HII regions may be as small as $<$0.03~pc, which
turns into an angular size $<1''$ at the typical distances of OB-type stars
($>$1~kpc). Such a small size, combined with the large opacity of the
free-free emission in the radio regime, makes it challenging to detect this
type of objects and even more to resolve them. The limited information available on the
earliest evolution of HII regions is frustrating, as this is considered a
crucial step in the formation of a high-mass star \citep{Keto_2003}.
The youngest, densest HII regions arise as the star contracts towards its
main sequence configuration. As long as accretion at high rates goes on, the
protostar may swell up \citep{Hosokawa_2010} and cool down, thus dramatically
reducing the ionizing photon output. This implies that the appearance of an
HC~HII region must correspond to the termination of the main accretion phase.
In this context, the morphology of the youngest HIIs tells us also about the
distribution of material immediately surrounding the forming massive star in
terms of the interaction between infall and outflow.

As a matter of fact, HII regions appear to remain in the most compact
phases of their evolution longer than expected on the basis of a simple
expansion model, as witnessed by the number of ultracompact HII regions in
the Galaxy \citep{Wood_1989, Mottram_2011}. This result implies the existence of some confinement,
which could be tightly related to the accretion mechanism itself, as shown by
\cite{Keto_2002}.  Alternatively, effects of density gradients and stellar winds may play a role 
\citep{artur_2006} or intrinsic variability driven by changes in the accretion rate \citep{DePree_2014}.
Therefore, shedding light on the earliest phases
of HII regions might be equivalent to unveiling the process of high-mass star
formation.

\begin{figure}[tbp]
\centering
\includegraphics[angle=-90,width=.45\textwidth]{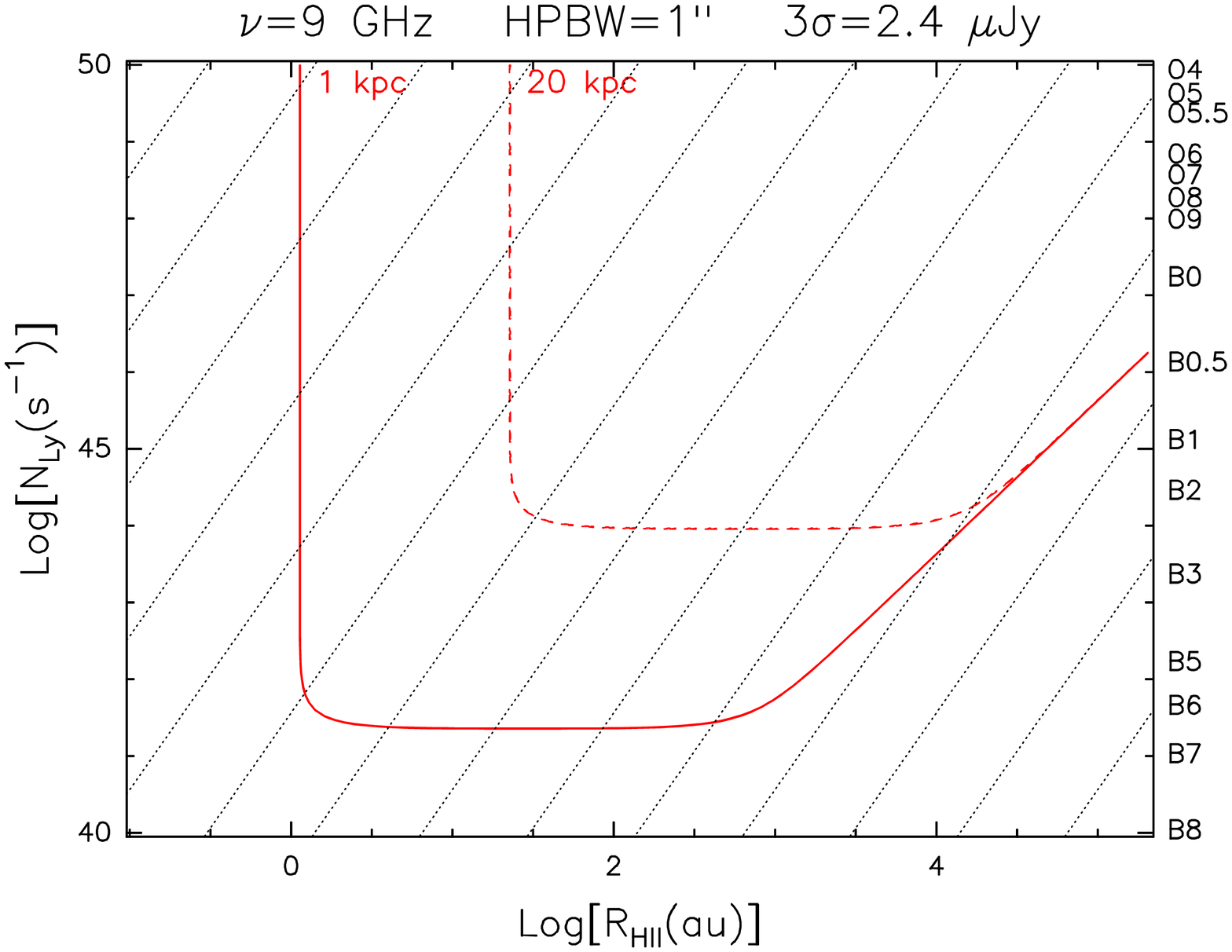}
\includegraphics[angle=-90,width=.45\textwidth]{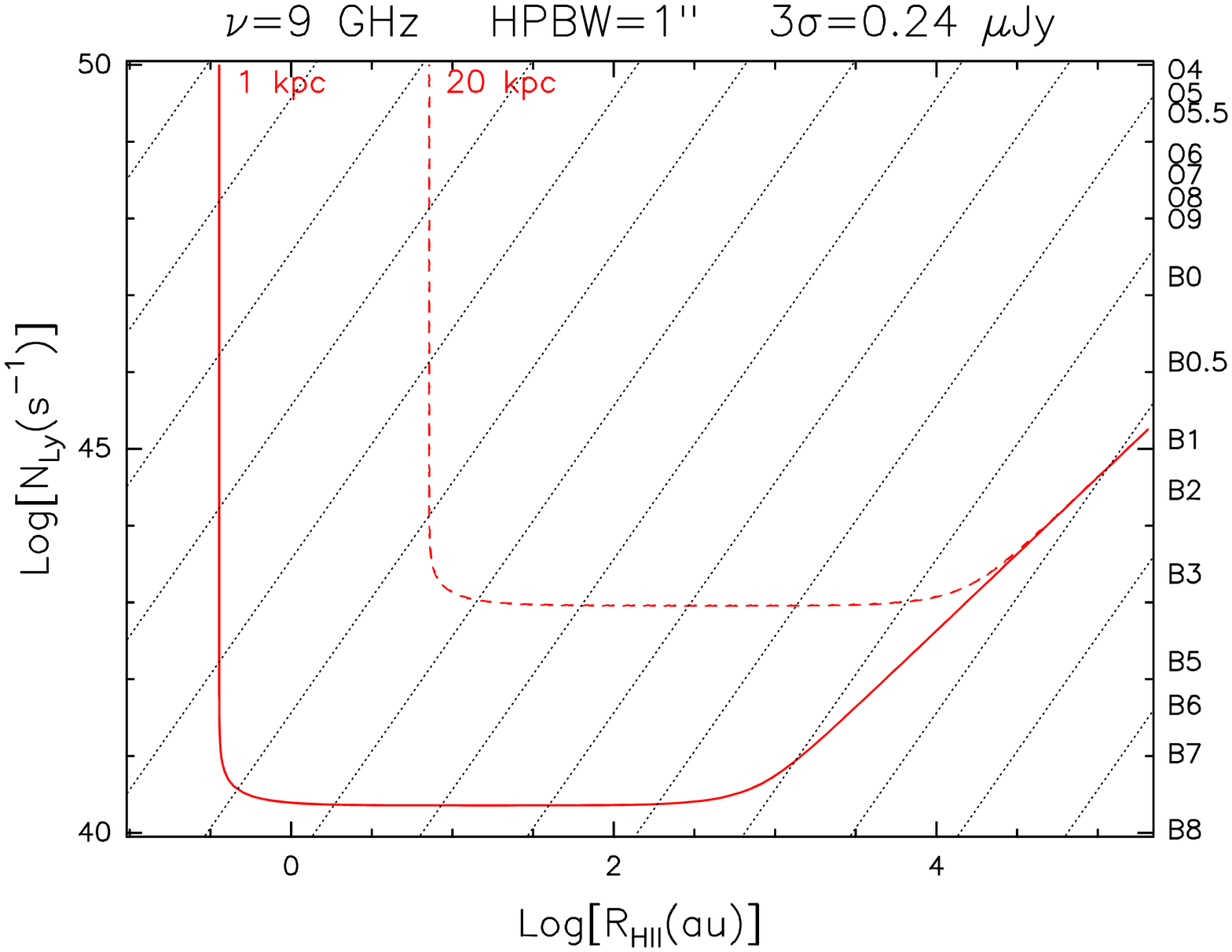}
\caption{{\bf Left:} Plot of the peak free-free continuum flux density of an homogenous,
isothermal (Str\"omgren) HII region as a function of the HII region radius
and Lyman continuum photon rate of the ionizing star. The calculation has
been performed assuming an observing frequency of 9~GHz and an angular
resolution of 1$''$. The two curves correspond to a 3$\sigma$ detection level
of 2.4~$\mu$Jy (obtainable in 10~min on-source with SKA1) at a distance of
1~kpc (solid curve) and 20~kpc (dashed). The dotted lines correspond to
fixed values of the HII region densities, ranging from $10^{-1}$~cm$^{-3}$
in the bottom right to $10^{12}$~cm$^{-3}$ in the top right, in steps of a
factor 10.
{\bf Right:} Same as left panel, for a detection threshold of 0.24~$\mu$Jy,
attainable with SKA in 10 minute  integration on-source.}
\hfill
\label{fska1}
\end{figure}

What is needed to boost our knowledge of (hypercompact) HII regions is both a
statistically complete sample and a detailed analysis of selected candidates.
Note also that radio observations should be better performed at higher
frequencies, as this regime is well suited  to actually determine the
morphology of the youngest, most compact HII regions.
SKA can fulfil all of these requirements, thanks to its superior sensitivity,
angular resolution, and frequency coverage. The sensitivity issue is illustrated in Fig.~\ref{fska1},
where we show the detection limits of SKA1 and SKA. In our estimate we have
assumed a classical (Str\"omgren) HII region ionized by a zero-age
main-sequence star. The figures plot the intensity of an HII region as a
function of the Str\"omgren radius and Lyman continuum of the ionizing star.
The curves correspond to the 3$\sigma$ level attainable with 10~min
integration on-source at 9~GHz with 1$''$ resolution, for a source distance
of 1~kpc (solid line) and 20~kpc (dashed). The HII regions falling above
the curve are detectable. One sees that even SKA1 will be sensitive enough
to detect HII regions as small as a few 10~au (the size of our Solar system!)
around a B2 star or earlier, across the whole Galaxy. The great potential of
SKA will permit to perform unbiased surveys of the Galactic plane in a
limited amount of time and with great sensitivity, and thus allow a complete
census of Galactic hypercompact (besides more extended) HII regions.

\begin{figure}[tbp]
\centering
\includegraphics[angle=-90,width=.65\textwidth]{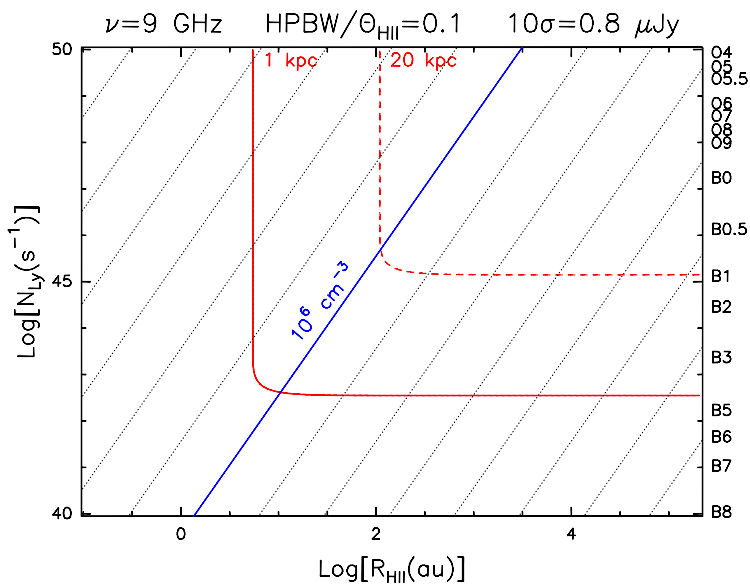}
\caption{Same as previous figure. The difference is that here we fix the
ratio between synthesized beam and source angular diameter diameter to 0.1
and allow for a S/N=10 for an integration of 10~min on-source with SKA
(resulting in 1$\sigma$$\simeq$0.08~$\mu$Jy). The blue line corresponds to an
HII region density of $10^6$~cm$^{-3}$.}
\hfill
\label{fres}
\end{figure}

Resolving a HC~HII region is more challenging. The maximum diameter is
$\sim$0.03~pc \citep{Kurtz_2005} (i.e. 0.6$''$ at a distance of,
e.g., 10~kpc) and requires a synthesized beam at least $\sim$10 times smaller
(i.e.  $<$0.06$''$) to be properly imaged. In Fig.~\ref{fres} we show a plot
analogous to those in Fig.~\ref{fska1}, but this time we fix the ratio
between the synthesized beam and the source angular diameter to 10. The
curves correspond to S/N=10 for 10~min integration on-source, assuming a
1$\sigma$ RMS of 0.08~$\mu$Jy for SKA at 9~GHz. Points above the solid blue
line have electron densities in excess of $10^6$~cm$^{-3}$, the minimum value
for a HC~HII region (see Kurtz 2005). We conclude that the SKA will resolve
HC~HII regions around B1 stars or earlier, all over the Galaxy.

In addition to the sensitivity and angular resolution, SKA will provide us
also with a broad instantaneous frequency coverage, thus allowing us to
measure the spectral index of the radio emission. Although not sufficient
by itself, knowledge of the spectral slope is a crucial piece of
information to discriminate between HII regions and thermal jets,
and determine the optical depth of the emission. Indeed,
radio jets might be confused with faint HII regions and are detectable even
at large distances (see Anglada et al.\ 2015).



\section{Late stages of stellar evolution}

SKA will also contribute to unveil the so-far missed populations of
Planetary Nebulae (PNe) and Supernova remnants (SNRs).  Those objects
represent the late stages of stellar evolution of low and intermediate mass
stars  and massive stars, respectively.  Details on the potential of SKA on
SNRs science can be found in Wang et al. (2015).  Low and
intermediate mass stars (LIMS: 0.8--8\,$\rm M_\odot$) constitute 90\%\ of
all stars which have died in the Universe. During their final evolution,
these stars eject between 40\%\ and 80\%\ of their total mass, enriched by
the products of nucleosynthesis. The ejecta are the source of half the
recycled gas in the Galaxy and are major contributors to chemical evolution
of the Galaxy. The mass loss also determines the final mass of stellar
remnants. The mass-loss process is one the main open problems in stellar
astrophysics. Current areas of research are: what drives the mass loss, how
is it affected by binary interactions and magnetic fields, and whether
there is a unique initial-final mass relation.
 
Planetary nebulae (PNe) are formed when the ejecta are briefly ionized by the
stellar remnant; they are known for their often beautiful morphologies.
Since they have a rich emission-line spectrum, they can also be used to trace
the kinematics of their host galaxies as well as serve as standard candles   \citep[e.g. ][]{Ciardullo_2010}.
Although they trace 90\%\ of all stars, they are  short-lived (a few 10$^4$ yr), hence relatively rare, with a
population density $\sim 10^{-6}$ of that of the total stellar population. This is
compensated by their high luminosity ($\sim 10^4\,\rm L_\odot$).

PNe are among the most numerous Galactic radio sources detected so far in
radio surveys (about 700 in the NVSS). However, our census of this type of
stars is far below the number expected from theoretical counts, which, on the
other hand, heavily depend on the assumptions on the previous evolution of the
stars. If, for example, only binary interaction is required to form a PN, we
can expect to observe $\sim$6600 PNe in our Galaxy, but this number can go up
to $\sim$46000, if binaries are not strictly necessary \citep{Jacoby_2010}.
The number of detected Galactic PNe is only $\sim$3000 \citep{Parker_2006}.

It is therefore possible  that there is a missing population of PNe. The
main reason for such a large mismatch between the expected and the observed
number of sources is that most PNe are hidden by dust absorption. Among other
factors that can hamper PN detections, there are the intrinsic low brightness
of the more evolved sources, which numerically dominate the PN luminosity
function and are then expected to constitute the main component of
volume-limited samples \citep{Parker_2006}, the interaction with the ISM,
which can disrupt the nebulae and shorten their lives, and the fact that
Galactic latitudes beyond $\pm$10 deg have not been adequately surveyed
\citep{Jacoby_2010}.


The sensitivity of the SKA and its ability to map large areas of the sky
quickly will allow us to detect every PN in the Milky Way. If we consider a
typical expansion velocity of 10 km~s$^{-1}$, the shell expands in
$2\times10^4$ yr to a distance of about 0.2 pc from the central star. Assuming
optically thin emission from an ionised mass of 0.1 M$_\odot$ ($n_e\;\sim$140
cm$^{-3}$) filling $2/3$ of the shell, we expect a flux density of about 4
$\mu Jy/beam$ at a distance of 61 kpc, over a beam of 0.8$''$. This means that
SKA1 will be able to detect and map evolved PNe not only in the whole Galaxy,
but also in the Magellanic Clouds.  The number counts, distribution and
brightness temperature will test the mass-loss models of LIMS. This survey can
be done at frequencies between 1 and 5\,GHz, with lower confusion and less
effect from optical depth (for the younger nebulae) at the higher frequencies.

Besides evolved PNe, the SKA will allow us to detect more objects in the rapid
transition from post-AGB stars to PNe, like CRL 618. This is a critical phase,
when the development and expansion of the ionisation front radically changes
the physical conditions in the circumstellar environment \citep{umana_2004, Cerrigone_2008, Cerrigone_2011}.
This transition lasts for a small fraction of the
post-AGB time (which amounts to some 10$^3$ yr), which makes it challenging to
detect stars going through it. Again, the deeper survey obtainable with the
SKA will find new transition stars by sampling a larger volume of our
Galaxy. In this context, it must be noted that transition objects are often
optically thick below 3 GHz, therefore high-frequency receivers will be
necessary to characterize the nebulae by observing their optically-thin
emission.  The flux may increase by 1\%\ per year or more, a variation readily detectable
if the survey is repeated over a few years.

Post-AGB stars such as CRL618 commonly show jet-like outflows. The origin and
driving is not clear but magnetic-driven models have recently come in vogue
\citep{Huarte_12}, strengthened by synchrotron emission from one
such jet \citep{Perez_13}. The SKA can detect synchrotron
components to the radio emission through its sensitivity and wide frequency
band: this will test the shaping mechanisms operating in PNe. This is best done at 1\,GHz.

Finally, the SKA can also measure the mass of the stellar remnants inside the
PN. This is done indirectly, through the evolution of the radio flux
density. The rate of heating of the central star (temperature increase in
K/yr) is a very strong function of the stellar mass. For stars which have not
yet reached their maximum temperature, the radio flux is expected to decrease
by 0.01-0.1\%\ per year, due to the decrease in number of ionizing photons
(hotter stars have fewer photons as each photon carries more energy). The rate
of change of the optically thin radio flux yields the stellar mass, to an
accuracy of better than 5\%.  This has so far only been done for NGC 7027
\citep{Zijlstra_2008}. The changes during the earlier evolution are faster
\citep{Cerrigone_2011}  but less deterministic. The SKA survey can detect a
0.1\%\ change for PNe as faint as 10mJy, which brings the Galactic Bulge
population in range. A deeper, targeted survey could go further.


These measurements will best be carried out at 4--6 GHz \citep{Zijlstra_1989, Aaquist_1990}.
 One way to
achieve the temporal sensitivity is to repeat the Galactic plane area of the
SKA survey  5--10 times, which would also improve sensitivity.


\section{Stellar Radio emission}
\subsection{General}
Stars emit, in the radio band, a negligible fraction of their total luminosity. In the case of the quiet Sun the ratio between its radio
and  its bolometric luminosity is less than $10^{-12}$. 
Nevertheless, in many cases, radio observations are the only way to reveal and study  astrophysical phenomena
that play a fundamental role in our understanding of stellar evolution.
This is the case  of stellar magnetic fields, whose topology can be  directly  mapped via  VLBI  observations \citep[e.g.][]{trigilio_1993, Peterson_2010} and its influence on energy amplification and subsequent release in stellar coronae can be investigated 
via the  correlation between radio and X-ray 
emission and the statistics  of stellar flares  \citep[and references therein]{benz_10};
of  stellar winds,  where  the detection of thermal radio emission  is widely used  to determine the mass-loss rate in hot stars \citep{Wright_975,Panagia_975, Scuderi_1998,Blomme_2011}
and has revealed to be particularly powerful when other diagnostics cannot be used,  as in the cases of dust enshrouded objects \citep{umana_2005}

 The improvement of the observational capabilities have lead to the
discovery of radio emission in  a wide  variety of stellar objects from all
stages of stellar evolution.  Broadly speaking, the brightest stellar radio
emission appears to be  associated with enhanced stellar mass-loss (winds,
nebulae) in the cases of   thermal emission (large emitting surface) or to
magnetically-induced phenomena, such as stellar flares, for  non-thermal
emission processes (high brightness temperature).  In the following we
briefly illustrate some  representative examples  for each type  of stellar
radio emitters.  \subsubsection{Thermal emitters} Thermal emission
(bremsstrahlung emission) is expected from winds associated with massive
stars, shells surrounding Novae and jets from symbiotic systems (O'Brien et
al.\ 2015), class 0 pre-main sequence (PMS) stars and
classical  TTauri \citep{White_2004}.

Massive stars play a fundamental role in the evolution of galaxies.
They are among major contributors to the interstellar UV radiation and, via their strong stellar winds,
provide enrichment of processed material (gas and dust) and mechanical energy to the interstellar medium.
Moreover, mass-loss from massive stars is very important for stellar evolution and understanding the different types of SN explosion.

Typical mass-loss rate for OB stars is of the order of $10^{-6} M_{\odot}/yr$, with  wind speeds of $(1-3)\times 10^{3} km/sec$.
More evolved massive stars include the classes of Luminous Blue Variables (LBVs) and Wolf-Rayet (WR).
Luminous Blue Variables are massive evolved stars in an highly unstable
state. Photometric and spectral variablities are as characteristic for that
phase as is a highly enhanced mass loss (up to few   $10^{-4}
M_{\odot}/yr$) in various wind phases in which the  wind velocities
alternate between slow and fast. WR are further evolved, hotter and more 
stable massive stars with similar strong mass loss. 
The strong  mass-loss of both object classes lead to 
extreme obscuration, in some cases, of the stellar surface. 
The mass-loss rate of massive star winds can be in principle derived from continuum radio observations, which trace the  ionized gas through its  optically thick free-free emission,  providing the distance to the source and 
the velocity of the wind are known.
A stellar wind has a typical spectral signature in the radio related to the radial density gradient of the wind  \citep{Panagia_975}.

Radio observations have been proven to be  more efficient  than usual diagnostics  for ionized gas, such 
as $H_{\alpha}$, because they don't suffer from extinction and are indeed the only way to probe ionized gas in very reddened objects, embedded in dense, dusty circumstellar material.
However,  while mass-loss rates from radio continuum observations have been  routinely  obtained from a large number of objects, quite recently it was realized that canonical mass-loss rates from massive stars need to be revised downwards, mostly because the assumed stellar-wind model does not include the effect of clumping or porosity  \citep{Blomme_2011}.

\subsubsection{Non-thermal emitters}
Much of our knowledge of non-thermal  emission from radio stars comes from the study 
of active stars and binary systems as a large fraction of them have been found to be strong radio 
sources  \citep{slee_87, drake_89, umana_91, umana_93, umana_98}.
Both classes of star are characterized by a magnetically heated outer atmosphere and  display all the manifestations 
of solar activity (spots, chromospheric active regions, coronal X-ray emission, flares).
In  binary systems the observed phenomenology   is  more extreme than in the solar case because of the 
forced rotation induced by tidal forces that contributes to generate a more  efficient dynamo action.
The radio emission arises from the interaction between the stellar magnetic field with mildly 
relativistic particles \citep[i. e. gyrosynchrotron emission]{Dulk_1985} and 
 is highly
variable. Two different regimes are usually observed: 
quiescent periods, during which a basal flux density of
a few mJy is observed, and active periods, characterized
by a continuous strong flaring which can last for several
days  \citep{umana_95}.

Non-thermal radio emission is also observed in  Ultracool dwarf (UCDs).
The class of UCDs  consists of stellar objects located on the boundary 
with sub-stellar bodies such as gas giant planets, including fully convective, very low mass M stars
(later than M6) and Brown Dwarf (BDs).

The recent discovery of intense radio emission pulses lasting a few minutes
from a number of UCDs has changed the conventional way to interpret their coronal emission physics;
such objects exhibit very low chromospheric H-alpha and coronal X-ray
activity and, consequently,  were expected to be radio quiet 
\citep[ and references therein]{mclean_12}. 
This detection has immense implications for our understanding of both stellar magnetic activity
and the dynamo mechanism generating magnetic fields in fully convective stars.
This manifestation of magnetic activity  is a significant departure from the incoherent
gyrosynchrotron emission model generally applied to cool stars and bears
 more resemblance to planetary auroral activity than coronal stellar activity,  indicating a possible transition in activity at the end of the main sequence. 
The radio pulses are thought to be due to highly beamed electron cyclotron maser emission and thus provide an accurate measurement of magnetic field 
strength at the location of the emission. This has been successfully used to provide the first measurements of magnetic field strengths for L and T dwarfs
\citep{berger_09, hallinan_07,Route_2012}. 

In several UCDs, such  coherent emission is periodic,
with a period consistent  with their respective rotational periods  \citep{berger_09, hallinan_07}. 
The long-term stability of the radio emission also indicates that the magnetic field
(and hence the dynamo) is stable over a long timescale.
How such fields are created and sustained remains a mystery.

Another example of non-thermal radio emitter is the class of Magnetic Chemically Peculiar stars (CPs).
They are B-A main sequence stars characterized by strong 
dipolar magnetic fields with axis tilted with respect to the rotational one (oblique rotator). 
There are no convective motions in their stellar envelopes,
and the magnetic field is thought to be fossil,  remnant of the dynamo fields generated in the pre main-sequence phase.

CPs can show radio continuum emission, variable with the rotational period, but stable in time  \citep{leone_93}.
The current model assumes that particles of the stellar wind are accelerated in the current sheets at Alfv\'en radius, and then
propagate in a thin magnetospheric layer toward the star, emitting gyrosynchrotron radiation  \citep{trigilio_04, leto_06}.
Radio observations at 1.4-2 GHz of the CP star CU Vir have revealed the presence of a coherent, highly directive, $100\%$ per cent
polarized radio emission component  \citep{trigilio_00},
interpreted as cyclotron maser.
In the framework of the radio emission from CP stars, maser amplification can occur in annular rings above the pole,
generating auroral radio emission similar to what has been detected in the planets of the solar system, including
the Earth \citep{trigilio_11}.
The coherent emission is stable on a time-scale of more than 10 years and has been used 
as a marker of the rotation of the star, revealing changes of the rotational period  \citep{trigilio_08}.
The well known topology of CPs magnetic fields  makes this type of object a
perfect template for studies of stellar magnetosphere, magnetoactive plasma, particle acceleration, and stellar spin-down.

\subsection{Radio stars detection forecast}
\begin{figure}[tbp]
 \centering
\includegraphics[width=.85\textwidth]{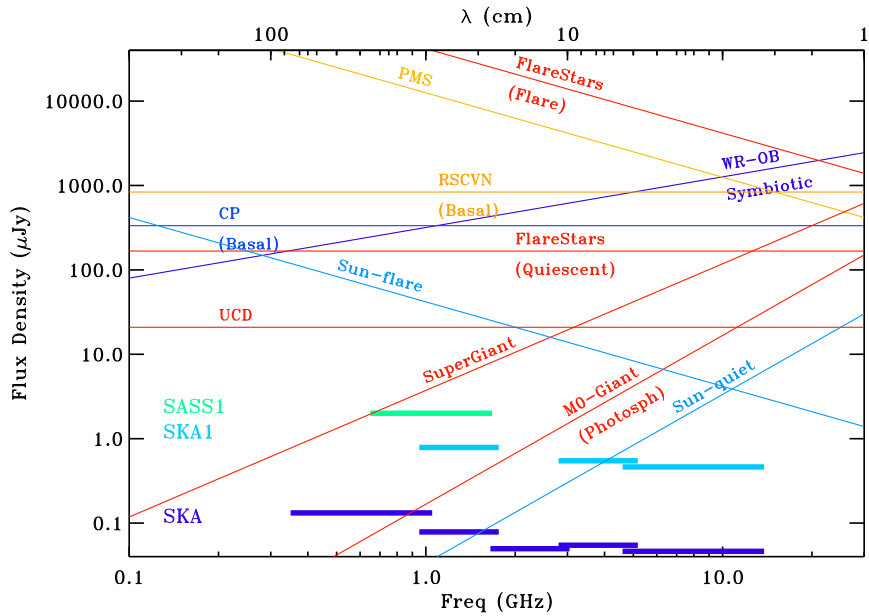}
\caption{Typical radio spectrum of several classes of radio emitting stars. Fluxes have been derived from the radio luminosity assuming a distance
as appropriate for each type of radio stars.  Detection limits for SKA1 and SKA have been computed for one hour integration time. See text for explanation.}
\hfill
\label{starska1}
\end{figure}

In the last years, a few hundreds of radio stars have been detected \citep{gudel_02} but   nearly all of the detections are the results 
of targeted  observations directed at small samples of stars thought to be likely radio emitters. 
Therefore, all the detections   suffer from a strong selection bias as the targeted observations were aimed  at  addressing  specific problems
related to some kind of peculiarities observed in other spectral regimes.
This approach has been proven to be quite productive but it is biased against discovering unknown, 
unexpected, or intrinsically rare objects, preventing a good knowledge of radio stars at the sub-mJy level.
In Fig.\,\ref{starska1}, schematic continuum radio spectra of several classes of radio emitting stars are drawn.
Fluxes have been derived assuming a typical radio luminosity 
\citep{seaquist_93, umana_93, gudel_02, berger_05, trigilio_08},
and a typical distance for each different types of radio stars: 10~pc for flare stars and late M-L, 100~pc for active binary systems, 1~kpc for supergiants, OB and WR, 500~pc for CP stars. The flux density of the quiet Sun has been derived from the solar quiescent radio luminosity assuming a distance of 10pc.
The spectral and sensitive characteristics of SASS1 and SKA-MID,  in both phases SKA1 and SKA, have been also drawn.
The sensitivity for SKA1 and SKA is obtained assuming
an integration time of 1 hour.  For SKA1, we assume that band 2, band 4 and band 5 will be available, while for the full SKA we assume the deployment of all five frequency bands.

Stellar winds and non-thermal radio emission from
many active binaries, flare stars and PMS stars will be easily
detected, within the considered distances,  with an rms sensitivity of $2\, \mu{\rm Jy} \,beam^{-1}$
(SASS1), even during its  early phase of deployment.
With SKA1 we  will be able to  detect a quiescent Sun, but only if band 4 or 5 is used.

\begin{figure}[tbp]
\centering
\includegraphics[angle=-0,width=.85\textwidth]{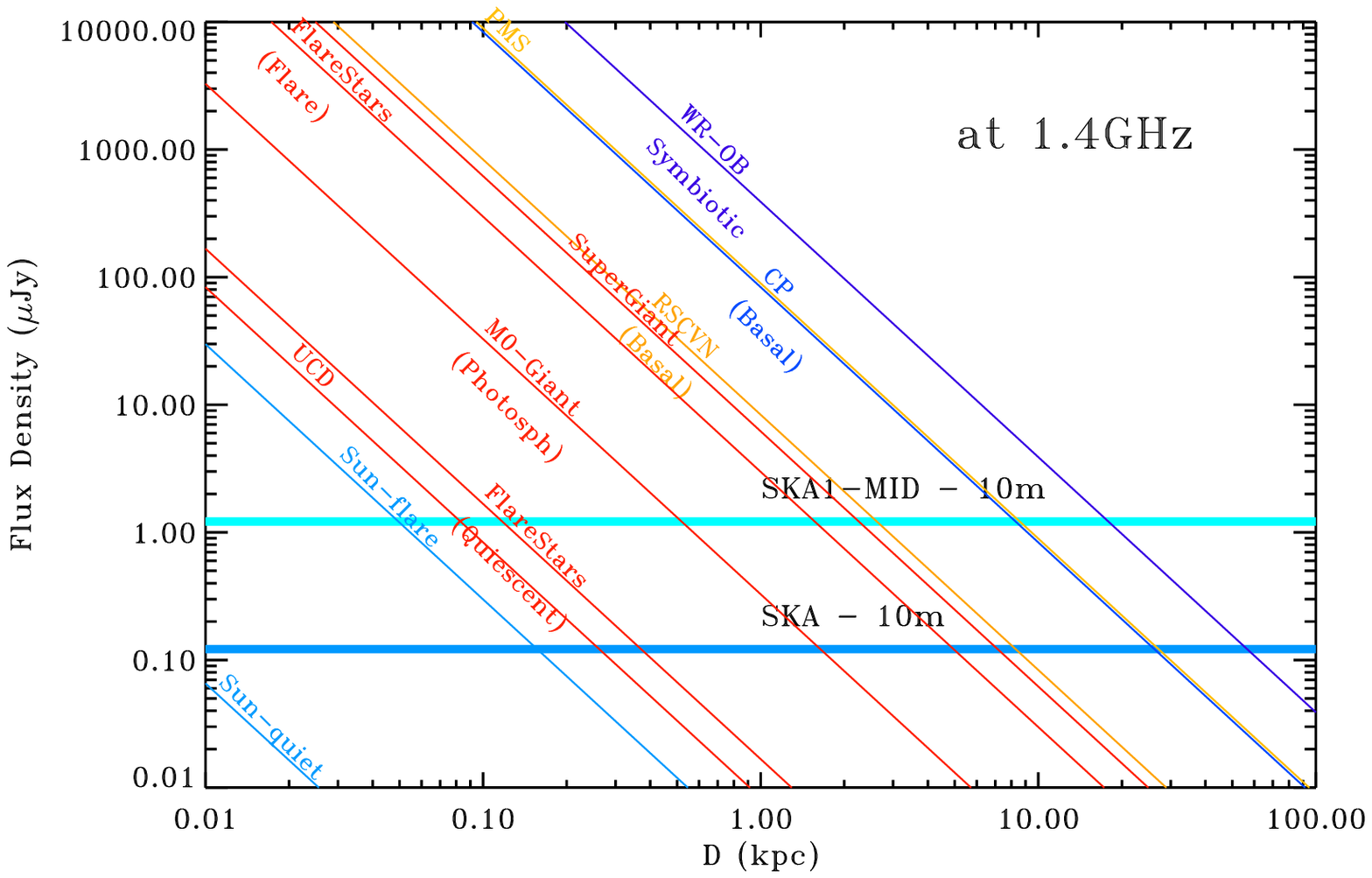}
\includegraphics[angle=-0,width=.85\textwidth]{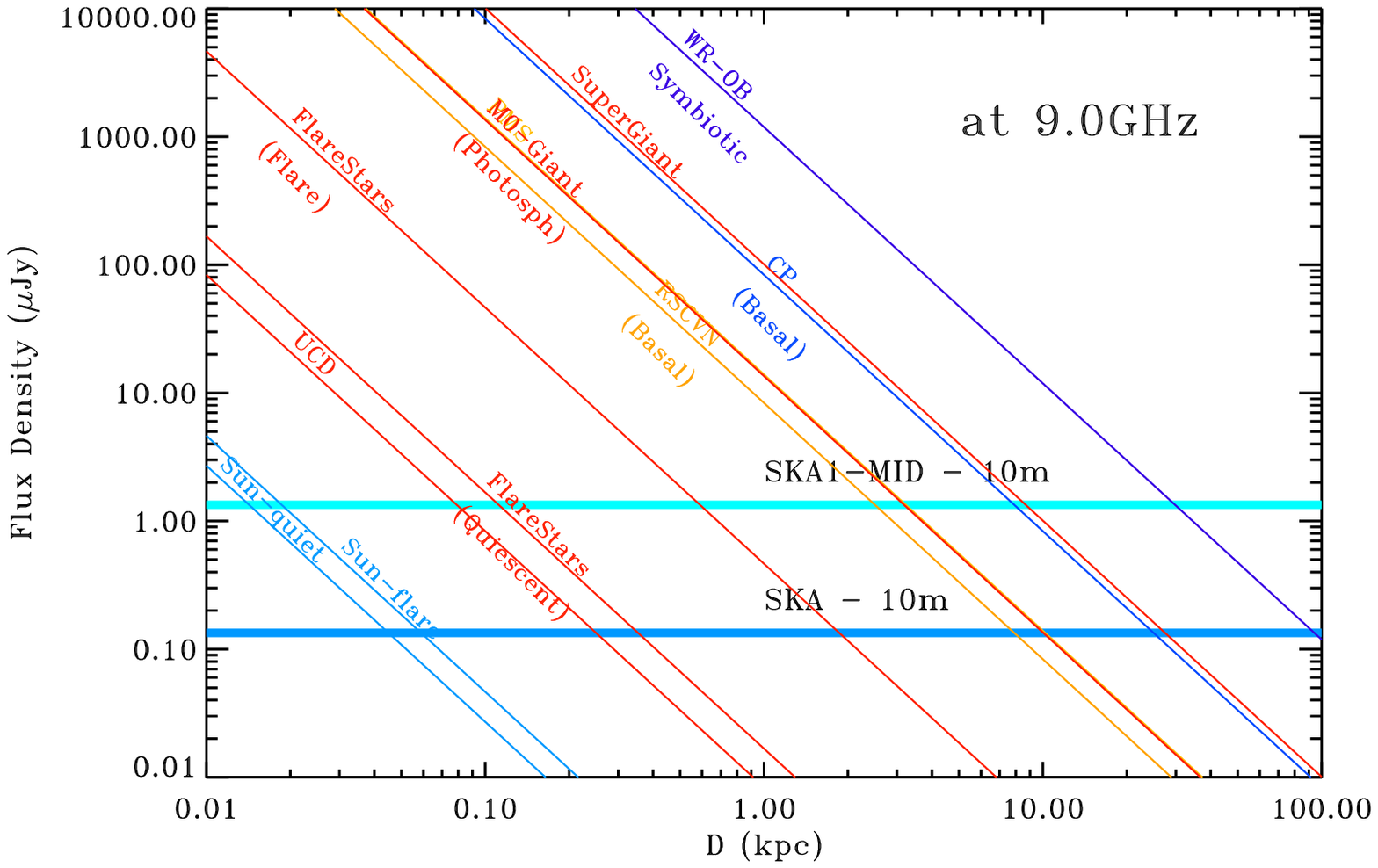}
\caption{{\bf Top panel:} Typical distances at which different  classes of radio stars can be detected, assuming as limit flux the sensitivity that can be achieved with 10 minutes
of integration band with SKA-MID in its band 2, in both SKA phase 1 and 2. {\bf Bottom panel}: Same as top panel  but for band 5.}
\label{starska2}
\end{figure}

Another way to see this is to estimate, starting from the typical radio luminosity and assuming a  limiting sensitivity, at  what distance is possible to detect a star belonging to a particular class of radio emitting object. 
With the  limiting sensitivity detectable flux density of SKA1-MID, for 10 minutes of integration time,   all the WR, OB and stars and Symbiotic systems of the Galaxy can be detected, while CP, PMS, RSCVn and Supergiants 
could be seen up to the distance of the Galactic Center (GC).
With the same integration time, SKA-MID will be able to detect almost all the above classes in the Milky Way, and probably in the nearby galaxies, providing sufficient angular resolution ($ \leq 0.02^{\prime \prime}$),
M Giant  photospheres to the distance of the GC, flare stars and UCDs within several hundreds pc, and "a quiescent Sun analog" up to 50~pc (Fig.\,\ref{starska2}).\\
The unique capabilities of SKA will allow to detect many classes of stars over the entire Milky Way. 
This will produce a real revolution in stellar physics as the radio properties of different stellar populations will be defined, allowing 
timely  comparisons with other stellar parameters, such as age, mass, magnetic fields, chemical composition, and evolutionary stages. \\
There are some particular  areas of stellar radio emission that will particularly
benefit from high sensitive  and high angular resolution  radio observations as those provided by the SKA.
These are  the study of non-thermal  stellar flares in active stars and
binary systems, the search for coherent events in different classes of stellar systems and the use of radio observations 
to derive mass-loss rate in massive stars.\\
It has not be established yet if the complete set of polarisation measurements
(Stokes I, Q, U and V) will be available with both SKA1 and SKA foreseen
observations. In particular circular polarisation information on a series of
non-thermal radio emitting stars will complete our understanding of particular
phenomena such as those related to coherent emission, allowing to immediately
point-out coherent radio flares occurring in a particular target. However, most
of the scientific goals, as reported in the following, could be reached even if
only I, Q and U maps will be available.

\subsubsection{Stellar Coronae: the solar-stellar connection:  }
In the Sun, during flares, 
beams of fast electrons travel down
magnetic fields and release their energy in the chromosphere,
where  heated plasma expands into coronal magnetic
flux tubes. This hot plasma then cools radiatively and
through conduction. Several attempts to use the solar model as a proxy 
 to understand the energetics and location of flares
in active stars have been carried out.  Various theoretical mechanisms have
been proposed, but more observations are clearly needed. 

Until now, all the information on the radio emission from active stars and stellar systems
originate from  targeted observations of very few, bright, well known objects, usually selected on the basis of the strong magnetic activity displayed in other spectral regions.

Even if limited to a very small sample, we have now some knowledge of   their radio emission characteristics 
(e.g. flare development, spectral and polarisation evolution, emission
mechanism, etc).  Deep radio measurements, as those that we can perform with SKA,  would significantly
enlarge the number of active stars and stellar systems, without bias effects, allowing new insights into
the physics of objects showing magnetic activity.
SKA1 would allow us to detect all the flares stars and active binary systems up to few kpc,  while SKA
will allow us to detect them over  the whole Galaxy (Fig.\,\ref{starska2}) and thus to increase of two orders of magnitude 
the number of active stars  that can be studied  in the radio.\\

The study of large samples of active stars will provide us with important clues on stellar magnetism and dynamo processes
as a function of internal structure of the stars and other physical parameters (mass, age, rotation...).
Moreover, a  clear understanding of the key parameters that control magnetic activity
in different types of stars is very important  since it affects habitability of possible orbiting planets. 

It will be possible to identify radio coronae across a range of cool type stars. 
Detailed studies of a large number of stellar coronae will improve our knowledge of   energy
release in the upper atmospheres of stars of different
mass and age. This will also permit to investigate the correlation between
radio and X-ray emission and thus study the occurrence of the 
 Neupert effect in stellar coronae \citep{gudel_09}.
 
Systematic, multi-epoch deep surveys  will enable 
the detection of serendipitous flaring activity and will hence permit
to derive the typical behaviour (occurrence rate, evolution, etc.) from a statistical study of
larger samples.
Multi-wavelength type follow-up observing campaigns would allow to study magnetic activity manifestations at different layers of the stellar  outer atmosphere
and to  explore the relationship of electron energization to the  long-lived centers of surface activity (photospheric spots), pointing out the existence of possible magnetic cycles also in the radio.

Finally, the unique sensitivity of SKA will allow to follow the development of flares with unprecedented  details and time-resolution.
A typical, solar-type  weak flare (0.1 mJy at 1.3 pc)  can be detected in 5 sec with SKA1 ($5 \sigma$).  Such kind of observation will provide 
new insights in the open question of coronal heating and, in particular, a piece of evidence that the radio quiescent corona is maintained by a series of micro-flares.

\subsubsection{Coherent events}

There is a growing evidence that  stellar radio flares can occur also as 
narrow band, rapid, intense and highly polarized  (up to 100\,\%) radio bursts, that are 
observed especially at low frequencies ($<$1.5 GHz). For their  extreme characteristics, they have been generally
interpreted as result of coherent emission mechanisms. such as the Electron Cyclotron Maser Emission (ECME).
Coherent burst emission has been observed in different classes of stellar objects: RS CVns and Flare stars
\citep{osten_08, slee_08}, Ultra Cool  dwarfs \citep{hallinan_08, Route_2012}
Chemically Peculiar stars \citep{trigilio_08, trigilio_11}, all having, as common ingredient,   
a strong and also, but not necessarily, variable magnetic field and a source of energetic particles. The 
number of stars where coherent emission has been detected is still limited to
few tens, because of the limited sensitivity of the available instruments and the stochastic nature of the events.

Deep radio observations, such those that SKA  will provide
the best opportunity to determine how
common coherent radio emission is from stars, stellar and
sub-stellar systems. The detection of coherent emission
in a large sample of different types of stars will have
immense implications for our understanding of both
stellar magnetic activity and the dynamo mechanism
generating magnetic fields in fully convective stars and
brown dwarfs \citep{hallinan_08, ravi_11}.

Coherent emission  observed  in binary systems and active stars and in UCD stars
shares several characteristics with that observed in CP
stars \citep{trigilio_00}, 
since both require a large scale
magnetosphere, and are similar to the low frequency
coherent radio emission observed in the planets with magnetic field of the solar system \citep{trigilio_11}.
To better understand the ECM in the wider context of plasma processes it is necessary to extend radio observations to a larger  sample of CP stars.
We want to stress here that CP stars  provide us with the unique possibility to study plasma processes in stable magnetic structures, whose topologies are 
quite often well determined by several independent diagnostics \citep{Bychkov_2005},  thus overcoming the problem of variability of the magnetic field  observed  in very active stars such dMe or close binary systems. 

The foreseen sensitivity of SKA, just in its first phase,  will allow to detect CP stars up to 10 kpc in 10 minute integration time (Fig.\, \ref{starska2}). 
Following \cite{Renson_2009}, we can assume the CP stars are uniformly distributed in space. This would imply that the number of radio detections 
will increase by about an order of magnitude, giving the opportunity to get a larger statistics of the physical conditions of the magnetospheres to be correlated  
to the ECM. 

If coherent emission is present in many radio active
stars, with the same characteristics, it will constitute
an excellent diagnostic for star magnetospheres, and
a powerful probe of magnetic field topology. 
In the case of UCDs the study of  ECM instability provides the only potential probe into magnetic field strengths for late-M, L
and T dwarfs.  

More observations of wider samples of active stars are also necessary to establish the percentage of active stars and binary systems that show coherent emission, 
exploring time-scales of its variability and how this is related to the basic physical parameters of the stars.
Follow-ups of the detected sample would allow to point-out any similarities between ECME from single stars and binary systems and thus discriminate between different
causes for the population inversion that drives the ECME events  \citep{slee_08}.


The discovery of other radio lighthouses similar to those  observed in  CU Vir  \citep{trigilio_08, ravi_11} will enable high precision
studies of the rotation period, and thus angular
momentum evolution, in different classes of stars.
 
\subsubsection{Mass-loss from Massive stars}

\begin{figure}[tbp]
\centering
\includegraphics[width=.75\textwidth]{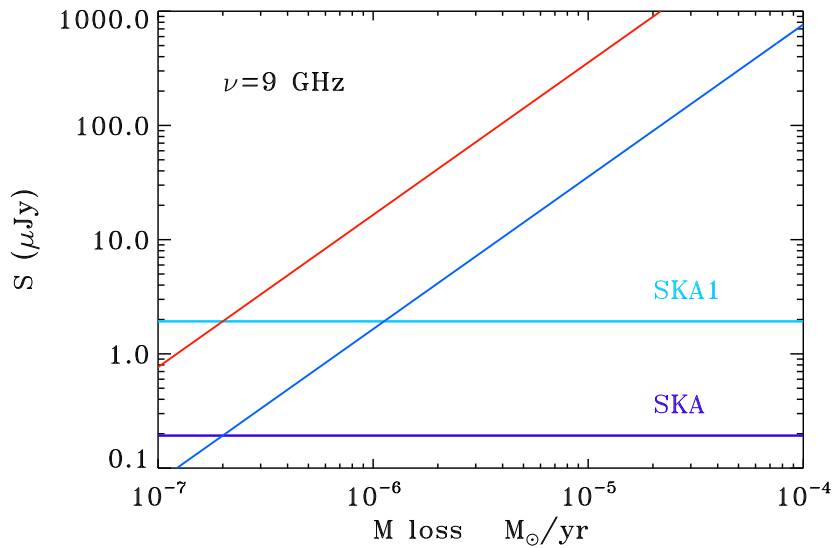}
\caption{Mass-loss rates detectable at the distance of the Galactic Center  (8\,kpc) for two values of wind velocities. 
Sensitivity (rms), for 10 minute integration time for both SKA1 and SKA  band 5, has been assumed.}.
\label{massloss}
\end{figure}

The recent discovery that the mass-loss rates derived for hot main sequence massive
stars need to be revised downwards (Blomme 2011), poses  serious questions on
the evolution of massive stars.  An  O-type star on the Main Sequence (MS), whose
mass may be as large as 150 $M_{\odot}$, will evolve into a Wolf-Rayet star,
with a typical mass not in excess of 30 $M_{\odot}$, but MS mass-loss rates are 
insufficient to account for such a huge mass-loss.  Severe mass-loss probably
occurs through strong stellar winds and/or eruptions during the post-MS
evolution.  Many luminous classes of stars belong to this phase, hot
supergiants (BSGs, B[e]s and LBVs), cool Yellow Hypergiants (YHGs) and  red
Supergiants (RSGs) stars.  The exact evolutionary path leading to a W-R, as
a function of the mass and rotation, is however not well constrained, 
mostly because a crucial piece of information,  i.e. mass-loss rates and 
lifetime and thus the total amount of mass lost during the post-MS  
is currently incomplete.

LBVs may play a key role in this scenario; as well as being characterized  by
a strong mass-loss, they  can also undergo giant eruptions, 
in which larger amounts of mass are being ejected. 
As a consequence of the strong stellar wind and/or the giant eruption,
circumstellar nebulae (LBVN) are formed (few $M_{\odot}$ for the wind
scenario, several $M_{\odot}$ by ejection), which are a few parsec in size
and have expansion velocities between 10-200 km/s, in extreme cases several
1000 km/s are detected \citep{Weis_2011}. The nebulae are seen in optical and 
IR emission lines, radio continuum emission and IR excess emission. 
The possibility that such spectacular mass-loss events
may be metallicity-independent  has greatly increased the interest in LBVs, as
this can have important implications for the mass-loss and therefore the
evolution of Population~{\small III} stars.
Observations of LBV eruption and other variabilities  of massive stars in 
local very low metallicity galaxies already hint at additional physics, 
which will improve the link from local massive 
stars to stars at the time of reionization \citep{Bomans_2011}. 

Mass-loss rates  from a number of LBVs and LBVs candidates have been recently
derived by radio observation in our own Galaxy \citep{umana_2005, umana_2010, umana_2012} 
and in the LMC \citep{Agliozzo_2012}.
 Radio observations have also allow to determine the morphology of the ionized
fraction of some LBVNs and to quantify the mass of ionized material  \citep{Buemi_2010, umana_2011}. 
When radio maps are combined with detailed
mid-IR maps, tracing the dusty component, the presence of multi-epoch
mass-loss events have been pointed out and estimates of the total content in mass
of LBVNs have been derived \citep{umana_2012, Agliozzo_2014}.
Of further interest will be a comparison of the radio data with optical/NIR 
images of several LBVN. \cite{Weis_2011} showed that the morphology of LBVN as 
deduced from 
[NII] or H$_{\alpha}$ images and a kinematic analysis \citep[i.e.]{Weis_2003}  is 
bipolar in about 50\%  of all nebulae;  for galactic LBVNs this rises to 75\%.
MIR dust nebulae appear to give different results, but are due to another 
driving mechanism, therefore radio observations are crucial for the
understanding of formation and evolution of nebulae, e.g. determining 
the influence of wind-wind interactions and stellar rotation.

In Fig\,\ref{massloss}, the mass-loss rates detectable at the distance of the
Galactic Center  (8~kpc), assuming the SKA1 and SKA sensitivity with a 10
minute  integration time, are shown.  The red  line refers to a wind
velocity  of $100\, km/sec$, while the blue  line to  $1000\, km/sec$. The minimum
detectable flux has been calculated for SKA-MID band 5, as it is at
higher frequency that we have the highest contribution from the optically
thick, thermal stellar wind.  Fig.\,\ref{massloss} indicates that with the
SKA, in both phases, we will  reach, in 10 minute  integration time,  a detection
limit  sufficient to measure also very small mass-loss rates ($\sim 10^{-7} \,
M_{\odot}yr^{-1} $) at the distance of the Galactic Center.  This would allow
studies, similar to those currently conducted on LBVs,  to be carried out
inside the  three massive stellar clusters, located near the Galactic Center
(Arches, Quintuplet and the Central Cluster).  These young stellar clusters are
more massive than any other cluster in the Milky Way and are likely to contain
massive stars at all stage of evolution,  including pre-main sequence, LBV and
W-R, allowing to explore a plethora of stellar winds and associated Nebulae
from a stellar population at the same age and, given the unique location,  to
study as these can be affected by environmental parameters.

\end{document}